\documentclass{article}

\usepackage[a4paper]{geometry}
\newgeometry{left=1.2in,bottom=1.2in,right=1.2in,top=1.2in}
\usepackage[table]{xcolor}
\usepackage{cite}
\usepackage{amsmath,amssymb,amsfonts}
\usepackage{graphicx}
\usepackage{textcomp}
\usepackage{listings}
\usepackage{import}
\usepackage{soul,color}
\usepackage[multiple]{footmisc}
\usepackage{xurl}
\usepackage{mathtools}
\usepackage{caption}
\usepackage{subcaption}
\usepackage{multirow}
\usepackage{rotating}
\usepackage{array}
\usepackage{tabularx}

\usepackage{hyperref}
\hypersetup{colorlinks,allcolors=black}

\begin{document}

\title{A Layered Architecture for Developing and Enhancing Capabilities in Large Language Model-based Software Systems
}

\author{
Dawen Zhang$^{1*}$, Xiwei Xu$^1$, Chen Wang$^1$, Zhenchang Xing$^1$, and Robert Mao$^2$
\\
$^1$CSIRO's Data61, $^2$ArcBlock
\\
$^*$David.Zhang@data61.csiro.au
}

\date{}
\maketitle

\begin{abstract}

Significant efforts has been made to expand the use of Large Language Models (LLMs) beyond basic language tasks. While the generalizability and versatility of LLMs have enabled widespread adoption, evolving demands in application development often exceed their native capabilities. Meeting these demands may involve a diverse set of methods, such as enhancing creativity through either inference temperature adjustments or creativity-provoking prompts. Selecting the right approach is critical, as different methods lead to trade-offs in engineering complexity, scalability, and operational costs. This paper introduces a layered architecture that organizes LLM software system development into distinct layers, each characterized by specific attributes. By aligning capabilities with these layers, the framework encourages the systematic implementation of capabilities in effective and efficient ways that ultimately supports desired functionalities and qualities. Through practical case studies, we illustrate the utility of the framework. This work offers developers actionable insights for selecting suitable technologies in LLM-based software system development, promoting robustness and scalability.
\\
\\
Keywords: Artificial intelligence, large language model, software architecture, agent
\end{abstract}

\section{Introduction}
\label{sec:intro}

The use of Large Language Models (LLMs) has expanded beyond traditional language-related tasks such as translation and question answering. This widespread adoption is largely driven by the generalizability and versatility of LLMs, which stem from being trained on vast amounts of diverse data sourced from the internet and human annotations, allowing them to capture patterns across many domains. Furthermore, LLMs use text as a flexible input/output interface, which makes interacting with them intuitive and adaptable to various contexts. Combined with advances in techniques that improve their ability to follow instructions and align with specific needs, LLMs have been increasingly applied to a variety of domain applications requiring flexibility and scalability, including software development~\cite{nijkamp2022codegen, hou2023large, austin2021program}, process automation~\cite{kim2024language, wang2024survey}, financial analysis~\cite{nie2024survey, wu2023bloomberggpt}, manufacturing~\cite{wang2023chatgpt, fan2024embodied}, education~\cite{dai2023reconceptualizing, sarsa2022automatic}, and scientific research~\cite{boiko2023autonomous, thirunavukarasu2023large, romera2024mathematical}.

However, despite their remarkable strengths, LLMs have clear limitations. For instance, even when trained on vast datasets, they often struggle with domain-specific knowledge and lack the specialized expertise needed for certain tasks~\cite{lewis2020retrieval}. Additionally, LLMs are prone to generating plausible-sounding but factually incorrect outputs~\cite{zhang2023siren}, commonly referred to as ``\textit{hallucination}.'' Since LLMs are primarily trained on the NLP task of next word prediction based on statistical probabilities~\cite{radford2018improving}, they may be less reliable and efficient than simpler but specialized tools for tasks that require a deeper understanding or internal task representation~\cite{zhou2024larger}. Furthermore, due to their language processing nature, they also interact with the external world exclusively through the interface of text, limiting effective access to the diverse interfaces of external tools or systems~\cite{yang2024swe}.

As LLMs are applied to more complex and varied tasks, new requirements continue to emerge. These novel demands often go beyond the initial capabilities of the models, poses significant challenges during the development of LLM-based applications. One of the most critical challenges lies in identifying the appropriate methods for implementing specific capabilities within LLMs~\cite{li2023chatgpt, tang2023struc}. Capabilities such as incorporating domain knowledge, enforcing specific constraints, or adhering to particular styles are not off-the-shelf features supported by LLMs, but can be achieved through various approaches. Incorporating domain knowledge is a particularly typical example, which can be achieved through techniques such as LLM fine-tuning or Retrieval Augmented Generation (RAG). Each method comes with its own trade-offs and characteristics, requiring careful design consideration. Importantly, choosing the wrong method for implementing a capability can lead to oversophisticated, inefficient, or ineffective solutions, and it may result in systems that are unreliable, fail to meet requirements, or underperform in production. Thus, selecting the right approaches for developing these capabilities is not merely a matter of preference but an architectural decision that can have a profound impact on the success of the application.

\begin{figure*}[h!]
\centering
\makebox[\textwidth][c]{\includegraphics[width=1.2\textwidth]{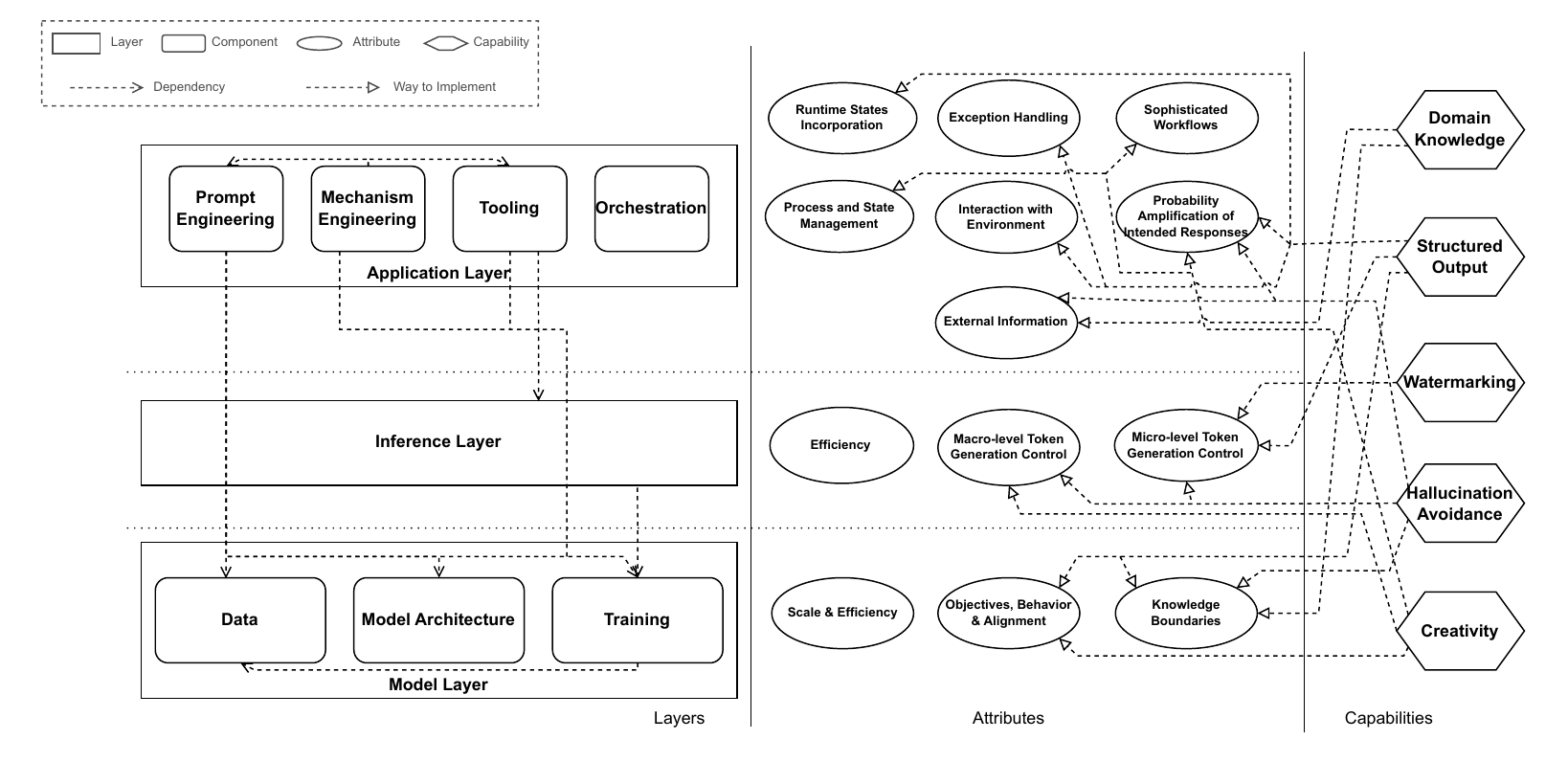}}
\caption{the conceptual architecture of the approach, including Components, Layers, Attributes, Capabilities, and their mappings}
\label{fig:architecture}
\end{figure*}

In this paper, we propose a layered approach for implementing capabilities in LLM-based applications by mapping them to the layers and components with corresponding attributes, shown in Fig~\ref{fig:architecture}. By aligning each capability with the relevant components at the correct layers, this approach reinforces essential software engineering principles, and ensures that capabilities are effectively implemented.

We illustrate the utility of this approach with several real-world example cases, showing how the right alignment between capabilities and implementation methods can lead to the appropriate solutions. Through these examples, we demonstrate how developers can use this approach to make informed design decisions.

This paper offers actionable insights into selecting the appropriate approaches for implementing capabilities, empowering developers to build effective and robust solutions for LLM-based applications.

The remainder of this paper is organized as follows.
Section~\ref{sec:motivation} explains the motivating scenario of this study, while Section~\ref{sec:framework} presents the layered approach, its underlying principles, and details of each layer. Section~\ref{sec:use-case} demonstrates the approach with various example cases, followed by 
a review of the existing literature related to this work
in Section~\ref{sec:related-work}. Section~\ref{sec:conclusion} concludes the paper.

\section{Motivating Scenarios}
\label{sec:motivation}

In this section, we present two motivating scenarios in which LLM-based application developers are likely to encounter the challenges addressed by our approach. While established solutions may exist for these scenarios, we aim to highlight the lack of a structured design process and conceptual framework. These hurdles, though partially addressed in some cases, continue to create gaps in other emerging scenarios that require careful design considerations.

\subsection{Application Scenario: Integrating Additional Data}

A key scenario in developing LLM-based applications is deciding how to integrate additional data, such as domain-specific knowledge. In general, this can be achieved through two main approaches: integrating the information as parameterized knowledge or integrating it as non-parameterized knowledge. Parameterized knowledge involves embedding the information directly into the model through training or fine-tuning, allowing the model to internalize the information. Non-parameterized knowledge supplies external information dynamically, typically through prompts, during inference, or post-generation, and is commonly referred to as Retrieval Augmented Generation (RAG)~\cite{lewis2020retrieval}.

When determining between parameterized knowledge and non-parameterized knowledge, while there are various factors that need to be considered, the primary difference between these two methods is whether the information is parameterized. In the case of parameterized knowledge, the model learns this information into its internal representations, making it readily available during inference. This brings a number of characteristics, for instance:

\begin{itemize}
    \item Minimal latency - As the information is already learned into parameters of LLMs, using these knowledge does not trigger retrieval from separate sources, and does not need information to be explicitly contained and processed as part of the input. This introduces minimal latency at inference time.
    \item Internal representations of information - Information is learned into the model's parameters, allowing it to form and store complex associations, relations and patterns within its internal representation. This integration enables the model to utilize the knowledge cohesively, drawing connections across tokens and adapting responses based on these embedded information.
    \item Resource-intensive at training - Integrating knowledge through training or fine-tuning can be resource-intensive, requiring significant labour, computational and time investment upfront.
\end{itemize}

In contrast, non-parameterized knowledge allows for flexible information integration at the inference stage by embedding relevant external information as context within the input provided to the large language model. This approach brings its own set of characteristics, such as:

\begin{itemize}
    \item Flexibility - Integrating information through non-parameterized approaches allows knowledge to be continuously added or updated independently of specific models and to be seamlessly transferred and utilized by any model without being tightly coupled to one.
    \item Additional input overheads - To supply external information, non-parameterized methods provide additional context to models within the input prompt, which consumes token space and may introduce computational overheads at inference.
    \item Surface utilization of information - Since the information is provided solely through prompts as external context and is not embedded within the model's parameters, it is processed based on the model's existing learned representations~\cite{xie2021explanation}. The model uses this information at a surface level, interpreting it in light of pre-existing associations.
\end{itemize}

These characteristics make each fitted to different sets of application scenarios of different nature. For example, if the developers want to develop the multilingual capability of the system by supplying data related to a new language to the model, parameterized method is the optimal choice, because language understanding demands the model to learn the representations that capture the language's semantic features and complex linguistic structures. Relying solely on in-context learning cannot establish the statistical relations and deep associations within the language, nor does it form robust internal representations, making it unlikely to produce coherent and sensible outputs.

\subsection{Application Scenario: Constraining Output Structure}

Another typical scenario is constraining the output structure of the language model. There are already relatively mature solutions, such as structured output based on constrained decoding, widely offered by commercial LLM vendors and the open-source community. However, these solutions were not available at the initial onset of demand, but instead, they went through an evolution process.

As LLMs gained popularity, users began to use them not only as chatbots but also to explore their potential as software components for executing tasks. These LLM-based components are connected with other components primarily through their text-based interfaces. However, since majority of non-LLM-based components may not be capable of processing natural language inputs, it becomes essential for LLMs to generate structured outputs that can be parsed and processed by other components. However, LLMs acquire their generalist capabilities during pre-training by learning the representations of vast unstructured textual knowledge from diverse sources in a probabilistic manner, which makes them natively incapable to produce stable structured outputs.

In order to connect LLMs with other components, developers attempted the challenge from various perspectives.
Initially, the most intuitive method of \textit{trial and error} was leveraged. Such attempt has led to sophisticated parsers in companion with carefully crafted prompts, but a structured output from LLMs was still not guaranteed. For instance, in the widely used LLM application framework LangChain, several structured output parsers, including \texttt{StructuredOutputParser}, \texttt{ListOutputParser} and \texttt{CommaSeparatedListOutputParser}, were introduced\footnote{Github Commit langchain-ai/langchain@df6c33d: \url{https://github.com/langchain-ai/langchain/commit/df6c33d4b31826650f5a04a5f9c352508bb4dd9d}}, to parse the outputs generated by the LLMs. The LLMs were expected to generate outputs based on text prompts that specify the desired output formats. A while after that, a new commit\footnote{Github Commit langchain-ai/langchain@ce5d97b: \url{https://github.com/langchain-ai/langchain/commit/ce5d97bcb3e263f6aa69da6c334e35e20bf4db11}} introduced additional trial-and-error-based output parsers, including \texttt{RetryOutputParser}, \texttt{RetryWithErrorOutputParser} and \texttt{OutputFixingParser}, 
to complement the structured output parsers. Since then, sophisticated retry, reflection, and fixing mechanisms have been integrated into the codebase, expanding it from initial 80 lines of code (LoC) to over a thousand LoC.

From a language model training perspective, generating outputs in specific formats has long been a traditional focus in the field. Fine-tuning has been an effective method for enforcing specific output formats in LLMs~\cite{dunn2022structured} by adapting the models with the datasets that exemplify the desired structures. 
As the demand for structured outputs increased, fine-tuning was offered by the commercial LLM vendors and the open source community explicitly as a means to reliable structured outputs, such as in OpenAI's GPT-3.5 Turbo fine-tuning API release.~\footnote{OpenAI GPT-3.5 Turbo fine-tuning and API updates (accessed 1 November 2024): \url{https://openai.com/index/gpt-3-5-turbo-fine-tuning-and-api-updates/}}. However, one significant challenge in this scenario is retaining the generalist capabilities of LLMs from the risk of catastrophic forgetting~\cite{kirkpatrick2017overcoming}. Furthermore, maintaining the flexibility of not requiring exhaustive fine-tuning across all possible output formats makes it more complex.

Another crucial stage where the output format can be enforced is the inference time. From a text generation perspective, the LLM decoding process selects a sequence of tokens based on the probability distribution for each subsequent token. By interpreting the output format schema as context-free grammars~\cite{parikh1966context}, the candidate token space can be constrained through logits manipulation, assigning the lowest probabilities to tokens that would violate the specified format~\cite{hokamp2017lexically}, a technique known as \textit{constrained decoding}. Since the probability of generated tokens is directly controlled by the formal representation of the expected output format, this approach ensures that only outputs strictly aligned with the specified format are generated. This method has been integrated into commercial models such as Claude 3\footnote{Claude 3 Model Family Announcement (accessed 1 November 2024): \url{https://www.anthropic.com/news/claude-3-family}} and GPT-4o\footnote{Introducing Structured Outputs in the API (accessed 1 November 2024): \url{https://openai.com/index/introducing-structured-outputs-in-the-api/}}, and open-source libraries like HuggingFace Transformers\footnote{Text Generation Inference Guidance (accessed 1 November 2024): \url{https://huggingface.co/docs/text-generation-inference/en/conceptual/guidance}} and llama.cpp\footnote{llama.cpp GBNF Guide (accessed 1 November 2024): \url{https://github.com/ggerganov/llama.cpp/tree/master/grammars}}. The core implementation of this method used in the HuggingFace Transformers library takes around 200 LoC\footnote{Github Commit dottxt-ai/outlines@18aaba1:\url{https://github.com/dottxt-ai/outlines/blob/18aaba1/outlines/processors/structured.py}}.

At the current stage of this evolution, LLM-based applications utilize a combination of three technologies, with inference-time constrained decoding positioned at the center for consistency and reliability. Training and fine-tuning, along with format-instructive prompts, assist the model in better understanding the schemas, ensuring that the outputs are both syntactically correct and semantically meaningful. Additionally, the retry and fixing parsers provides an extra layer of assurance that the outputs are of expected formats.

\subsection{Motivation and Scope}

From these two scenarios, we observe that implementing a specific capability within LLM-based applications can lead to various technological paths, with the suitability of each technology dependent on the specific requirements of the scenario.
Identifying these technological paths may require significant engineering efforts, and incur time and resources cost on it.
Moreover, these technological paths may operate at different layers. For example, in the constraining output structure scenario, fine-tuning targets the knowledge learned into the model parameters, while constrained decoding focuses on the token selection process, and prompt-based approaches address non-parameterized inputs to the model.
Although these different approaches can be employed to implement the same capability, they may occur at distinct decoupled layers of a LLM-based application, leading to significantly different expectations, outcomes, and implications for developers. Consequently, it is essential for developers to possess a systematic understanding of the software architecture of LLM-based applications and to reinforce software engineering principles to make informed architectural decisions.

Therefore, the motivation behind this work is
to propose an
approach
which serves to map anticipated capabilities to a potential technology stack for implementation, even in the absence of specific existing or known solutions.
\section{The Approach}
\label{sec:framework}

\subsection{Glossary}

To enhance clarity for the reader, we define two key terms used in this approach.

\paragraph{Capability} ``\textit{Capability}'' is a widely used term in LLM research, but there lacks the explicit and rigorous conception of the term~\cite{anwar2024foundational}. In this study, we define it as \textit{``the ability to effectively perform a category of tasks or exhibit a specific type of behavior,''} which is consistent with current LLM studies~\cite{anwar2024foundational, wei2022emergent, wang2024survey, shevlane2023model, zhou2023far} as well as its philosophical and anthropological roots.

\paragraph{Attribute} ``\textit{Attribute}'' in this approach is defined as \textit{``inherently being responsible for certain quality, feature, or operation of the system,''} refined from its dictionary definition of ``a quality or feature regarded as a characteristic or inherent part of someone or something.''

\subsection{Principles of the Approach}

\textit{``It is the mapping of a system's functionality onto software structures that determines the architecture's support for qualities.''}
\begin{flushright}
--- Software Architecture in Practice~\cite{bass2012software}
\end{flushright}

We extend this principle to the software architecture for LLM-based application by \textbf{mapping capabilities onto the appropriate layers and components with corresponding attributes.}

Layers and components are defined following the principle of separation of concerns. We present the layers, components and their attributes with additional properties in Section~\ref{sec:framework:layers}. Using this architecture, we illustrate the process of how to map desired capabilities onto the appropriate structures in Section~\ref{sec:framework:cap-mapping}.

\subsection{Layers}
\label{sec:framework:layers}

Inspired by layered software architectures such as n-tier architecture~\cite{richards2015software} and layered Internet protocol stack~\cite{rfc1180}, we break an LLM-based application into layers based on the nature of relevant development activities, including \textit{Model Layer}, \textit{Inference Layer}, and \textit{Application Layer}. For each layer in the architecture, we examine the nature of the components and their roles in supporting specific attributes. The overall structure is shown in Fig~\ref{fig:architecture}.

\subsubsection{Model Layer}
The Model Layer involves the foundational aspects of building an LLM, from the data collection to the architecture design and training process. The output of this layer is the large language model, which is a machine learning model typically built upon transformer or a variant architecture trained using data on natural language processing (NLP) tasks such as causal language modeling (next-token prediction). Through this process, \textbf{the model learns a statistical representation of language from the data, and embeds in its parameters.} We divide the Model Layer into three core components, i.e. \textit{Data}, \textit{Model Architecture}, and \textit{Training}.

\paragraph{Data} Data used for training and fine-tuning LLMs goes through a process of selection, collection, and preprocessing, serving as the source from which LLMs learn the representations. In state-of-the-art (SoTA) LLMs, training data is typically collected from diverse sources, including publicly available internet content, inputs from human data workers, knowledge obtained from domain experts, user interactions and feedback, and distilled data from the output of LLMs~\cite{khan2022subjects, biswas2022art, hsieh2023distilling}.

The quality and scope of the training data directly determine what information is embedded within the internal representation of LLMs and thus subsequently set the boundaries of model's knowledge and capabilities. Beyond learning factual knowledge from the training data, LLMs also capture the patterns, relationships, and biases present in the training data, all of which contribute to shaping the model's behaviors and responses. Moreover, these boundaries, behaviors, and responses are not fixed, as LLMs can be further adapted and refined through continual pre-training or fine-tuning, where additional data is used to update the model's internal representations.

\paragraph{Model Architecture}

The Model Architecture defines the structure and underlying mechanisms of LLMs. The vast majority of LLMs are built on the transformer architecture, which leverage a self-attention mechanism that allows the model to capture long-range dependencies and contextual relationships between tokens in a sequence~\cite{vaswani2017attention}. This makes transformers effective for a wide range of NLP tasks, from translation and summarization to text generation.

The capability of model is largely tied to the scale of model, often measured by the number of parameters. The scale of model is a critical factor that may lead to the emergence of model abilities~\cite{wei2022emergent}. Larger models have shown emergent abilities that are absent in smaller models, such as few-shot learning. These abilities are likely a result of the higher number of parameters leading to increased capacity to capture complex patterns and relationships from the training data~\cite{xie2021explanation}.

However, this comes with trade-offs in terms of computational cost and resource requirements. Larger models demand significant computational resources at training and inference, including more performant computing units and larger memory, impacting the latency and cost of the model. To address these challenges, some LLMs utilize a Mixture of Experts (MoE) architecture~\cite{shazeer2016outrageously, zhou2022mixture}, which offers a balance between the scale and efficiency. MoE models consist of multiple expert sub-networks, with only a subset of experts activated for each input, reducing the computational overhead of the inference. In contrast, smaller models, though less capable, are faster and more efficient, making them more feasible to be used for real-time applications or deployed on edge devices.

\paragraph{Training} Training is the process where model learns from the data and embeds what it learns into parameters. While the dominant approach remains next-token prediction, various training objectives and fine-tuning techniques have been developed to address specific needs and enhance performance across different applications.

Next-token prediction is a widely adopted training objective for LLMs, where the model learns to predict the next token in a sequence by estimating the probability distribution of tokens based on the preceding context. This task, also known as causal language modeling, aligns well with autoregressive models including GPT~\cite{radfordimproving}, allowing model to generate fluent and coherent text. Another text generation task is masked language modeling, adopted by BERT~\cite{devlin2019bert}, trains model to fill in the random blanks within a sequence of tokens. This bidirectional context, where the model makes use of both preceding and following tokens, enables its ability to understand the full sentence structure and semantic relationships but at the same time also makes it less fit for sequential text generation.

If certain information or patterns are absent in the pre-training dataset, the model can be fine-tuned on task-specific data to refine its knowledge base and behavior. As the demand for using LLMs for specific tasks has increased~\cite{radford2019language}, methods like instruction tuning, direct preference optimization (DPO), and reinforcement learning from human feedback (RLHF) have become popular to align models with human preferences and improve their ability to follow instructions~\cite{zhang2023instruction, rafailov2024direct, ouyang2022training}. These models are fine-tuned on crafted dataset or reward functions so that the patterns and human preferences are captured into parameters of models. These fune-tuning methods can also be used for specialized purposes, such as setting rules for LLMs~\cite{bai2022constitutional} or enabling models to autonomously call tools~\cite{schick2024toolformer}. The objective of next-token prediction can be further adapted by combining with techniques like zero-shot learning, chain-of-thought (CoT)~\cite{wei2022chain}, Monte-Carlo tree search (MCTS)~\cite{grill2020monte} and Reinforcement Learning. For instance, Quiet-STaR~\cite{zelikman2024quietstar} enhances reasoning by guiding models to ``think before speaking'' using intermediate rationale tokens before predicting next token to improve capabilities for reasoning tasks.

Open-source LLMs as well as commercial LLMs provide the interfaces for fine-tuning. For example, both OpenAI\footnote{Fine-tuning now available for GPT-4o: \url{https://openai.com/index/gpt-4o-fine-tuning/}} and Anthropic\footnote{Fine-tune Claude 3 Haiku in Amazon Bedrock: \url{https://www.anthropic.com/news/fine-tune-claude-3-haiku}} provide fine-tuning for their most advanced LLMs to application developers. HuggingFace's transformer library also supports both full fine-tuning and parameter efficient fine-tuning (PEFT)~\cite{hu2021lora} over all available models. This provide application developers the access to the model layer to customize the model based on their requirements.

\paragraph{Summary}

The Model Layer is responsible for shaping the internal representations of LLMs. The Data component defines the boundaries of the model's knowledge and capabilities, response patterns, and relationships from the training data into the model's parameters. The Model Architecture component determines the structural design, which influence both the scale and efficiency of the model and subsequently the emergence of abilities. Finally, the Training component dictates model's text generation mechanisms, refining model behavior and incorporating additional knowledge, alignment and capabilities through training objectives and fine-tuning techniques. These attributes with their corresponding components and developer access levels are shown in Table~\ref{table:model-layer}.

\begin{table}[h!]
\centering
\begin{center}
\begin{tabularx}{\columnwidth} { | m{0.235\columnwidth} | m{0.18\columnwidth} | m{0.235\columnwidth} | m{0.235\columnwidth} | }
 \hline
 \textbf{Attribute} & \textbf{Component} & \textbf{Dev Access (Commercial)} & \textbf{Dev Access (Open Source)} \\
 \hline
 Knowledge Boundaries & Data \& Training & Fine-tuning & Full Customization \\ 
 \hline
 Scale \& Efficiency & Model Architecture & Pre-set Options & Full Customization  \\
 \hline
 Objectives, Behavior \& Alignment & Data \& Training & Fine-tuning & Full Customization \\
 \hline
\end{tabularx}
\end{center}
\caption{The attributes determined by the Model Layer with corresponding components and levels of developer access.}
\label{table:model-layer}
\end{table}

\subsubsection{Inference Layer} The Inference Layer utilizes the model's learned representations to generate responses, a process known as decoding. In transformer architecture, the attention mechanism is involved to determine the most relevant context when forming the probability distribution of the next token~\cite{vaswani2017attention}. In each decoding step, the model produces a probability distribution over all tokens. This distribution is generated by applying the softmax function to the logits derived from the model's final hidden state, and the resulting probability distribution serves as the basis for selecting the next token in the sequence.

The key objective in inference layer is \textbf{deciding how tokens are chosen from the logits}. The na\"ive way is to select the token with the highest probability, known as \textit{greedy sampling}. While efficient, greedy sampling makes the generated sequence too rigid and deterministic. To make the output diverse, methods like top-k~\cite{fan2018hierarchical} and top-p~\cite{holtzman2019curious} samplings consider a subset of candidate tokens based on their ranks or cumulative probabilities. Beam search-based decoding uses a heuristic approach of sampling multiple candidate sequences and select the tokens based on the overall probability of the sequence~\cite{freitag2017beam}. Applied on the logits before produce probability distribution with softmax, temperature scaling manipulate the logits by scaling it with a temperature value, so that lower temperatures make the probability distribution sharper, leading to more deterministic outputs, while higher temperatures flatten the distribution, making the selection more stochastic. These strategies represent macro-level approaches to token selection.

As the inference layer controls the overall process of token generation, it can also exercise micro-level control, managing the generation in a fine-grained manner. This involves influencing not just the global token selection strategies, but also making more crafted adjustments during the decoding process to steer the output. As decoding involves processing the logits, logits manipulation makes it possible to directly boost or suppress certain tokens, enforce specific semantic patterns across the sequence, or promote or eliminate the generation of particular tokens or sequences. This also enables \textit{constrained decoding}~\cite{hokamp2017lexically}, where the probability of tokens violating predefined constraints is set to zero, ensuring that the generated output strictly adheres to the desired structure or format.

In addition to controlling the tokens generated, a primary concern in the Inference Layer is how to generate tokens efficiently. Efficient decoding focuses on minimizing latency, reducing computational overhead, and optimizing resource usage while maintaining the quality of generated outputs. One way to achieve that is speculative decoding, which uses a smaller assistant model alongside the main LLM during inference~\cite{leviathan2023fast, chen2023accelerating, li2024eagle}. The assistant model quickly drafts preliminary candidate tokens, which the primary model can then accept or correct based on the logits from a single forward pass.
Other methods may involve lower level customization, including paging the K-V cache~\cite{kwon2023efficient}, efficient attention mechanism~\cite{dao2022flashattention}, precision reduction~\cite{kumar2024scaling}, and parallelism~\cite{aminabadi2022deepspeed}.

Open-source community support a wide range of Inference Layer features through various libraries, including HuggingFace Text Generation Inference\footnote{Text Generation Inference: \url{https://huggingface.co/docs/text-generation-inference/}} and llama.cpp\footnote{llama.cpp - LLM Inference in C/C++: \url{https://github.com/ggerganov/llama.cpp}}. These tools offer developers extensive customization over inference settings, covering macro-level strategies, micro-level controls, and efficiency optimizations. In contrast, commercial LLMs typically provide limited customization options for the Inference Layer. Inference efficiency is centrally managed by the vendors, which is a reasonable encapsulation for simplified interfacing and user experience. The exception is prompt caching, which incurs lower cost when using repeated prefix sequences in their queries. For macro-level control, vendors usually provide basic parameter adjustments for temperature, top-k, and top-p sampling, but more advanced sampling methods often remain black-boxed. At the micro-level, customization is primarily restricted to structured output options via format schemas, offering basic control over output format while lacking support for more fine-grained token-level control.

\paragraph{Summary}

The Inference Layer is responsible for the text generation procedure of LLMs. Macro-level strategies steer the global configurations of output, while micro-level controls craft the details of the generated sequence in fine-grained manner. The inference efficiency aspect optimizes latency of the inference, through approaches such as assisted generation, KV cache operations, efficient attention mechanisms, precision reduction and parallelism. These perspectives with respective developer access levels are shown in Table~\ref{table:inference-layer}.

\begin{table}[h!]
\centering
\begin{center}
\begin{tabularx}{\columnwidth} { | m{0.22\columnwidth} | m{0.4\columnwidth} | m{0.295\columnwidth} | }
 \hline
 \textbf{Attribute} & \textbf{Developer Access (Commercial)} & \textbf{Developer Access (Open Source)} \\
 \hline
 Macro-level Token Generation Control & Temperature, Top-K \& Top-P & Full Customization \\
 \hline
 Micro-level Token Generation Control & Structured Output & Full Customization \\
 \hline
 Efficiency & Prompt Caching & Full Customization  \\
 \hline
\end{tabularx}
\end{center}
\caption{The attributes determined by the Inference Layer with their levels of developer access.}
\label{table:inference-layer}
\end{table}

\subsubsection{Application Layer}

The Application Layer translates LLM's text generation power into functionalities, \textbf{bridging the gap between the raw input/output text generation paradigm of LLMs and fully functional applications}. From a conceptual viewpoint, the Application Layer centers on four key components: \textit{Prompt Engineering}, \textit{Mechanism Engineering}, \textit{Tooling}, and \textit{Orchestration}. While additional components exist in LLM-based applications, such as user interface (UI) and application hosting, these aspects closely align with traditional software development practices and are not unique to LLM-based applications.

\paragraph{Prompt Engineering}

Prompt Engineering focuses on designing prompts that guide the LLM to produce the desired outputs. The practice of Prompt Engineering is rooted in the foundations established by the Model Layer, where the model learns its internal representations from training data and is often fine-tuned to follow instructions. However, the probabilistic nature of LLMs remains unchanged. The generated responses are inherently influenced by the learned statistical patterns and the underlying probability distributions. The core principle of Prompt Engineering is to steer the model's probability distribution, amplifying the likelihood of generating the intended response~\cite{liu2023pre}.

One effective strategy in Prompt Engineering is to reduce the ambiguity of the query. Vague query descriptions tend to produce inconsistent outputs, as they are less likely to align with the specific patterns the model has learned during training.
By providing clear and explicit instructions, the model's focus can be narrowed, guiding it toward a more concentrated area of its learned representation.

Another effective strategy is to supply additional guiding tokens directly within the prompt or to trigger the model to generate them before providing the final response. This helps steer the probability distribution towards the desired outcome by establishing a clearer context or guiding the model's reasoning process~\cite{xie2021explanation}. The approach leverages patterns learned during training, where the model has encountered sequences involving intermediate reasoning steps, explanations, and structured processes.
Since the inference process considers all previously generated tokens, this method enhances the likelihood of producing structurally coherent, contextually relevant, and logically consistent outputs, better aligning with specific task requirements.

Examples of these methods include zero-shot learning methods such as \textit{role playing}~\cite{white2023prompt}, \textit{step-by-step prompting}~\cite{kojima2022large}, and \textit{explain-then-answer prompting}~\cite{shwartz2020unsupervised}, and few-shot learning methods such as the vanilla few-shot prompting~\cite{brown2020language} and chain-of-thought prompting~\cite{wei2022chain}. Additionally, prompt optimization can be achieved through manual or automated tuning methods~\cite{liu2023pre}.

\paragraph{Mechanism Engineering}
Mechanism Engineering~\cite{wang2024survey} involves integrating LLM's text generation into modular mechanisms, that employ structured approaches that guide the module's outputs towards achieving broader objectives or solving complex problems. Mechanism Engineering designs abstract processes that leverage the model's generative abilities in a systematic way. These mechanisms typically consist of multiple individual text generation processes, each serving orthogonal or complementary objectives aimed at reaching a specific goal. Different generation processes are often connected through symbolic or rule-based programs, represented using certain structures, such as directed acyclic graph (DAG).

While LLMs generate tokens in an autoregressive and linear manner, this generation process inherently lacks explicit representations for iteration, recursion, branching, and conditionals. Mechanism Engineering addresses this limitation by providing an external representation that helps maintain state across different stages of text generation. By integrating symbolic structures and logical controls, it enables the LLM to incorporate external information or runtime state, and perform sophisticated workflows.

Retrieval-augmented generation (RAG)~\cite{lewis2020retrieval} inject external knowledge into the prompt to provide additional context for knowledge-intensive tasks. The mechanism has evolved by allowing LLM to actively query external information, such as Agentic RAG~\cite{jiang2023active} and Retrieval-Interleaved Generation (RIG)~\cite{radhakrishnan2024knowing}.

The trial-and-error method involves repeatedly incorporating error information into the input to the LLM, enabling it to improve its responses until the task is successfully accomplished.~\cite{wang2024survey}. The ReAct framework~\cite{yao2023react} implements an interleaved reasoning and acting mechanism, allowing the model to incrementally execute tasks by alternating between thought processes and actions. Reflexion~\cite{shinn2024reflexion} introduces more sophisticated mechanisms that enable the model to iteratively work on a task, incorporating information from both its trajectory and experience to enhance performance over time.

Self-consistency~\cite{wang2022self} and tree of thoughts (ToT)~\cite{yao2024tree} both adopt multi-path methods to explore the solution space of the task. In self-consistency, multiple paths are generated for a given prompt, and the final solution is determined by the voting mechanism, which selects the most consistent or frequent answer among the generated responses from all paths. In contrast, Tree of Thoughts (ToT) employs the LLM to evaluate intermediate steps along the path.

\paragraph{Tooling}

Without tools, LLMs are only generators of text, lacking the capacity to perform automation tasks, validate and iteratively refine proposed solutions, or access external information. Since LLMs can only interact with the environment via text-based input and output, they require tools to be integrated and adapted for use through natural language. The LLM itself cannot directly interact with tool interfaces, so it instead relies on intermediary components that interpret the generated text to invoke external tools.

In passive tool calling, such as in vanilla Retrieval-Augmented Generation (RAG)~\cite{lewis2020retrieval}, the decision to call a tool (i.e., retriever for the case of RAG) is made externally, not by the LLM itself. The result of the tool call is then supplied back to the LLM as part of the prompt context. This approach does not require the LLM to be aware of the existence of tools or understand the interfaces for using them.

In active tool calling, the LLM is given a degree of agency, allowing it to decide when and which tools to invoke based on the context of the task. The model can generate structured text (e.g., JSON) that follows specific conventions designed to trigger tool calls directly, or it can describe the tool invocation in natural language. In scenarios where the LLM uses unstructured natural language to specify the tool call, the responsibility for interpreting and executing the tool request is delegated to a specifically fine-tuned LLM, for instance the ToolFormer~\cite{schick2024toolformer} and DataGemma~\cite{radhakrishnan2024knowing}. This setup allows the primary LLM to maintain flexibility in how it interacts with tools while offloading the responsibility of explicitly conducting tool calls to a specialized component.

\paragraph{Orchestration}

The orchestration component manages the chaining of LLMs, integrates tools into the application, and maintains the states and overall process of the application.

LLM chaining involves linking together multiple LLM calls to form a cohesive process, tackling tasks in a divide-and-conquer manner. Each call builds upon the responses from previous ones, allowing the system to break down complex tasks into smaller, manageable sub-tasks. This enables the application to handle multi-step problems effectively.

The orchestration component also determines which tools are included in the application, how they are utilized, and how they are shared across multiple LLM calls. Different tools may have distinct execution behaviors, such as asynchronous processing or requiring specific input formats, which the orchestration component must accommodate.

States of the application as well as both short-term and long-term memory are maintained as part of orchestration. It tracks session data, user configurations, previous interactions, and the sequence of LLM calls, ensuring that context is preserved throughout the entire task execution. Additionally, the orchestration component handles errors and exceptions, particularly when the LLM generates unexpected outputs. It includes mechanisms to detect issues, trigger corrective actions, or adjust the process, ensuring a smooth and consistent execution.

\paragraph{Summary}

The Application Layer is responsible for leveraging the text generation power to building functional applications. The Prompt Engineering component focuses on crafting effective prompts that amplify the probability of generating desired responses. The Mechanism Engineering component enables sophisticated workflow operations and supports the incorporation of runtime state, allowing for advanced processes beyond simple text generation. The Tooling component extend the abilities of LLMs by integrating external tools, enabling the model to perform automation tasks, validate solutions, and access external information. Finally, the Orchestration component manages the coordination of the LLM's text generation, external tools, and user inputs, ensuring a smooth flow of information and the robust execution of workflows. These attributes with their corresponding components are shown in Table~\ref{table:application-layer}.

\begin{table}[h!]
\centering
\begin{center}
\begin{tabularx}{\columnwidth} { | m{0.471\columnwidth} | m{0.471\columnwidth} | }
 \hline
 \textbf{Attribute} & \textbf{Component} \\
 \hline
 Probability Amplification of Intended Responses & Prompt Engineering \\
 \hline
 Runtime States Incorporation & Mechanism Engineering \\
 \hline
 Sophisticated Workflows & Mechanism Engineering \\
 \hline
 External Information & Mechanism Engineering \& Tooling \\
 \hline
 Interaction with Environment & Mechanism Engineering \& Tooling \\
 \hline
 Exception Handling & Mechanism Engineering \& Orchestration \\
 \hline
 Process and State Management & Orchestration \\
 \hline
\end{tabularx}
\end{center}
\caption{The attributes determined by the Application Layer with corresponding components.}
\label{table:application-layer}
\end{table}

\subsubsection{Intra-layer and Inter-layer Dependencies}

The layers and components of the framework are not isolated. There are dependencies across and within layers,
such that achieving the desired capability may require support from other components within the same layer, i.e., Intra-layer Dependencies, or from components across different layers, i.e., Inter-layer Dependencies.

\paragraph{Intra-layer Dependencies}

In the Model Layer, components often depend on one another to work effectively. For instance, certain fine-tuning methods rely heavily on the availability and format of specific data. To perform instruction tuning, the data must be curated in the form of instruction-output pairs. Without this structured data format, the fine-tuning process cannot align the model effectively tune the model to follow instructions.

In the Application Layer, Mechanism Engineering has dependencies on both Prompt Engineering and Tooling components. Mechanism Engineering often requires carefully crafted prompts to guide the LLM towards desired responses. Additionally, certain mechanisms rely on the integration of external tools to complete the task.

\paragraph{Inter-layer Dependencies}

The Inference Layer often relies on components from the Model Layer to achieve optimal performance and efficiency. For example, efficient inference methods, such as running the model in reduced precision, may depend on specific training methods like quantization-aware training conducted in the Model Layer.

The Application Layer depends heavily on both the Model and Inference Layers. For instance, Prompt Engineering is influenced by the model architecture, particularly the scale of the model, which determines its ability to perform tasks like few-shot learning. Additionally, the presence of specific representations learned from data is crucial, as effective prompts rely on amplifying the model's probability distribution to guide responses toward desired outputs.
The effectiveness of certain prompts or mechanisms, especially those relying on instruction-following, often requires instruction tuning at the Model Layer to align the model's behavior.
Moreover, the transferability of prompts across different models depends on characteristics of models from the Model Layer~\cite{liu2023pre}.

In Application Layer, to build tools effectively, especially when employing active and direct tool calling, 
the support for structured output generation in the Inference Layer is beneficial, as it helps the LLM produce responses in the correct format required by the tool interfaces.

\subsection{Mapping Capabilities onto Layers}
\label{sec:framework:cap-mapping}

The Capability Mapping aims at aligning specific system capabilities with the attributes and components across the layers of the architecture. We will use the capability of generating JSON output as an example to illustrate this process.

\paragraph{Attributes Identification}

The first step in implementing a capability is to identify its relevant attributes across the layers. As demonstrated above, different layers are responsible for distinct attributes of LLM-based applications. Each capability may correspond to one or more layers, depending on its characteristics and requirements. Additionally, some attributes may become depreciating in necessity due to the presence of other attributes.

The capability of generating JSON output has several key attributes, shown in Table~\ref{table:cap-attr-mapping}.

\begin{table}[h!]
\centering
\begin{center}
\begin{tabularx}{\columnwidth} { | m{0.235\columnwidth} | m{0.15\columnwidth} | m{0.53\columnwidth} | }
 \hline
 \textbf{Attribute} & \textbf{Layer} & \textbf{Explanation} \\
 \hline
 Knowledge Boundaries & Model Layer & Having the knowledge of JSON syntax and structure in the representation of model is helpful for generating compliant output \\
 \hline
 Objectives, Behavior \& Alignment & Model Layer & Having model to follow explicit instructions improves the ability to generate output in a specific format \\
 \hline
 Micro-level Token Generation Control & Inference Layer & Fine-grained control at the time of token generation can ensure that the output adheres to the JSON format \\
 \hline
 Probability Amplification of Intended Responses & App Layer & Using clear and explicit prompt examples can increase the likelihood of generating the response in desired structure \\
 \hline
 \rowcolor{lightgray}
 External Information & App Layer & If the model's internal knowledge does not sufficiently cover JSON structure, external information can be supplied to the LLM \\
 \hline
 Interaction with Environment & App Layer & Integrating external tools can validate the response format and provide feedback for correction \\
 \hline
 \rowcolor{lightgray}
 Exception Handling & App Layer & Error detection and correction mechanisms handle cases where LLM fail to generate desired output format \\
 \hline
\end{tabularx}
\end{center}
\caption{Attributes related to the capability of generating JSON output. Light gray rows are identified as depreciating.}
\label{table:cap-attr-mapping}
\end{table}

Upon reviewing these attributes, we identify several depreciating instances. The knowledge about JSON syntax within Knowledge Boundaries (Model Layer) may reduce the necessity for External Information (Application Layer) if model's training data already includes sufficient examples of JSON structures. Moreover, Micro-level Token Generation Control (Inference Layer) can effectively enforce JSON format, making extensive Exception Handling (Application Layer) less critical. In light of this, both can be implemented in lightweight manner. By resolving depreciating instances, we avoid redundancy and oversophistication while maintaining effectiveness.

\paragraph{Solution Architecture}

Based on the identified attributes, we can design a solution architecture that aligns solutions with the appropriate components in each layer. It is important to account for Intra-layer and Inter-layer Dependencies. A simplified description for the JSON generation example is shown in Table~\ref{table:cap-component-mapping}.

\begin{table}[h!]
\centering
\begin{center}
\begin{tabularx}{\columnwidth} { | m{0.1\columnwidth} | m{0.24\columnwidth} | m{0.575\columnwidth} | }
 \hline
 \textbf{Layer} & \textbf{Component} & \textbf{Solution} \\
 \hline
 \multirow{3}{1em}{Model Layer} & Data & Include examples of instructions and JSON documents during fine-tuning to help model learn the instructions, syntax and patterns \\
 \cline{2-3}
 & Model Architecture & Ensure model selected is capable of few-shot learning \\
 \cline{2-3}
 & Training & Apply instruction tuning or fine-tuning on JSON generation tasks \\
 \hline
 Inference Layer & \multicolumn{2}{m{0.75\columnwidth}|}{Manipulate logits to prevent the generation of format-violating tokens at each decoding step} \\
 \hline
 \multirow{4}{1em}{App Layer} & Prompt Engineering & Ensure the clarity of prompt, and provide few-shot examples for generation \\
 \cline{2-3}
 & Mechanism Engineering & Implement lightweight exception handling mechanism \\
 \cline{2-3}
 & Tooling & Integrate JSON parser \\
 \cline{2-3}
 & Orchestration & Ensure that the workflow can detect and handle exceptions \\
 \hline
\end{tabularx}
\end{center}
\caption{The solution at each layer for implementing the capability of generating JSON output.}
\label{table:cap-component-mapping}
\end{table}

\paragraph{Access Resolution}

In cases where certain components in the architecture are gated or restricted, alternative solutions need to be considered.

In the JSON output generation example, if the Inference Layer is gated, developers can consider utilizing open-source alternatives. Alternatively, developers can compensate the gated components by enabling fallback mechanisms, such as the exception handling previously considered depreciating, which can be enabled to validate and correct the output format with mechanisms such as trial-and-error.

However, developers must also consider the cost-effectiveness of implementing such sophisticated and depreciating workarounds, especially if there is a strong likelihood that these capabilities will be centrally provided by vendors in future updates, driven by high demand. In these cases, it is important for developers to evaluate their own priorities to decide whether to implement these solutions or wait for vendor support, based on their demands and resource constraints.

\paragraph{Evaluation}

While the capability mapping provides a useful framework for guiding the implementation, it cannot replace the need for thorough evaluation of the solutions. It is crucial to conduct comprehensive testing, including ablation studies and continuous monitoring, to assess the effectiveness of each component. By dynamically adjusting the sophistication of each component in response to evaluation feedback, developers can ensure both the effectiveness and efficiency of the implementation.
\section{Use Case Evaluation}
\label{sec:use-case}

In addition to the JSON output generation example, we evaluate the usefulness of the framework with a variety of use cases.

\paragraph{Creativity}

Creativity refers to the capability of generating original ideas. While LLMs cannot truly generalize beyond the data they have been trained on, they can produce responses that appear creative by leveraging different aspects across the layers, as demonstrated in Table~\ref{table:case:creativity}.

\begin{table}[h!]
\centering
\begin{center}
\begin{tabularx}{\columnwidth} { | m{0.24\columnwidth} | m{0.15\columnwidth} | m{0.525\columnwidth} | }
 \hline
 \textbf{Attribute} & \textbf{Layer} & \textbf{Explanation} \\
 \hline
 Knowledge Boundaries & Model Layer & A broad and diverse training dataset equips the LLM with a vast pool of knowledge, enabling it to generate ideas that may be previously unseen by users. The model can also create new combinations by linking concepts or knowledge from different domains. \\
 \hline
 Objectives, Behavior, \& Alignment & Model Layer & Less control and alignment over human preferences may lead to more stochastic and unconventional responses. \\
 \hline 
 Macro-level Token Generation Control & Inference Layer & Decoding strategies such as temperature adjustment and top-k may lead to less deterministic output. \\
 \hline
 Probability Amplification of Intended Responses & App Layer & Crafted prompts reinforce the model to align responses with the forms of creativity it has learned during training. \\
 \hline
 Sophisticated Workflows & App Layer & Workflows and mechanisms that mimic human brainstorming and idea selection process to generate sensible ideas. \\
 \hline
\end{tabularx}
\end{center}
\caption{The solution at each layer for provoking creativity.}
\label{table:case:creativity}
\end{table}

Depend on the type of creativity needed, different solutions across layers can be selected or combined for desired effect. For example, Microsoft Copilot uses the temperature parameter to control the level of creativity in responses\footnote{Microsoft Copilot for Technical Leaders: \url{https://learn.microsoft.com/copilot/tutorials/learn-microsoft-copilot?tutorial-step=1}}. In contrast, for more specialized applications like scientific research agents, sophisticated workflows are employed for controlled creativity in creative tasks~\cite{huang2023benchmarking}. However, such complex workflows may be overly resource-intensive for general applications, where users only require a degree of semantic variation, rather than exhaustive creative exploration.

\paragraph{Call Caching}

Call Caching refers to the capability of reusing results from previous LLM calls to reduce computational costs and overheads.
The system can solutions to achieve it, as shown in Table~\ref{table:case:caching}.

\begin{table}[h!]
\centering
\begin{center}
\begin{tabularx}{\columnwidth} { | m{0.18\columnwidth} | m{0.15\columnwidth} | m{0.585\columnwidth} | }
 \hline
 \textbf{Attribute} & \textbf{Layer} & \textbf{Explanation} \\
 \hline
 Efficiency & Inference Layer & By storing and reusing attention states for repeated long prefix sequences, the system can skip recomputing these prefix sequences. \\
 \hline
 Process and State Management & App Layer & Hash-based lookup for storing and retrieving previous input and output pairs can avoid the same query being recomputed. \\
 \hline
\end{tabularx}
\end{center}
\caption{The solution at each layer for call caching.}
\label{table:case:caching}
\end{table}

LLM vendors including OpenAI and Anthropic integrate prompt caching at Inference Layer, which significantly reduces inference costs for bulk queries with same long prefix.

LangChain supports caching for tool calls in Application Layer, which is beneficial when LLMs are wrapped behind these tools, as it prevents redundant computations. Similarly, AIGNE\footnote{AIGNE: \url{https://www.aigne.io/}} provides an abstraction where components, including LLMs, are treated as blocklets, which are callable by users or other blocklets. Caching interactions between blocklets helps avoid repetitive nested invocations. These Application Layer caching mechanisms may not be effective in scenarios where the prompts vary across requests, but they can be useful in cases involving high-computational components like LLMs, where the same request is repeated multiple times within a period. In such cases, caching can minimize cost and overhead.

\paragraph{Long Context}

ChatGPT was only accepting 4096 tokens at the initial released and was insufficient for advanced usage such as codebase analysis. A larger context window enables the model to capture rich context or handle long documents. The solution at each layer is shown in Table~\ref{table:case:context}.

\begin{table}[h!]
\centering
\begin{center}
\begin{tabularx}{\columnwidth} { | m{0.155\columnwidth} | m{0.14\columnwidth} | m{0.62\columnwidth} | }
 \hline
 \textbf{Attribute} & \textbf{Layer} & \textbf{Explanation} \\
 \hline
 Scale \& Efficiency & Model Layer & Employ specialized attention mechanisms like Efficient Attention~\cite{shen2021efficient} and extended positional encodings such as RoPE~\cite{su2024roformer} to handle long input sequences. \\
 \hline
 Sophisticated Workflows & App Layer & Shortening the prompt through prompt compression mechanisms at Application Layer such as LLMLingua~\cite{jiang2023llmlingua} allow more content to fit within the model's context window. \\
 \hline
\end{tabularx}
\end{center}
\caption{The solution at each layer for long context.}
\label{table:case:context}
\end{table}

While the Model Layer solutions are native and robust for implementation, it may be unavailable for commercial models if long context is not supported. Solutions at the Application Layer, such as prompt compression, offer a flexible workaround, but they typically result in higher inference costs, and may be less robust and more prone to information loss.
\section{Related Work}
\label{sec:related-work}

Several studies have examined the design and architectural considerations of LLM-based systems. Zhou et al.~\cite{zhou2024taxonomy} present a decision model for foundation model-based agents, analyzing trade-offs between various options for each type of architectural element. Lu et al.~\cite{lu2024towards} propose a layered reference architecture for foundation model-based systems, emphasizing a pattern-oriented approach that focuses on component orchestration. Liu et al.~\cite{liu2024agent} compile a comprehensive design patterns catalogue for foundation model-based agents.

In specific areas, Shamsujjoha et al.~\cite{shamsujjoha2024taxonomy} provide a taxonomy focused on runtime guardrails for LLM agents, exploring a set of guardrail options at each level of system. Wang et al.~\cite{wang2024survey} introduce the concept of Mechanism Engineering, surveying relevant methods from the perspective of design mechanisms and applications of autonomous agent. Liu et al.~\cite{liu2023pre} offer an extensive survey on various prompting strategies for LLMs as well as their relevant enabling methods required during pre-training and inference. Welleck et al.~\cite{welleck2024decoding} explore inference-time algorithms, highlighting various objectives of decoding strategies. Zhang et al.~\cite{zhang2023instruction} provide a detailed survey on instruction tuning, discussing aspects including dataset preparation, tuning methods, and efficient tuning techniques.

While these works provide valuable insights into specific aspects, problems, or a particular layer of LLM-based applications, they address isolated components without a unified perspective. In contrast, our work fills this gap by offering a holistic layered approach that systematically examines the characteristics of each layer within LLM-based applications and guide the development of capabilities across these layers.

\section{Conclusion}
\label{sec:conclusion}

This paper presents a layered approach for implementing capabilities in LLM-based applications, decoupling the system into three distinct layers: Model Layer, Inference Layer, and Application Layer. Each layer is assigned specific attributes based on the characteristics of layers and their components. By aligning capabilities with relevant attributes across layers, this approach enables developers to systematically identify implementation strategies.

The proposed framework was evaluated against several typical use cases, demonstrating its usefulness and feasibility in guiding the design of LLM-based applications. Our work bridges the gap between high-level capability requirements and the underlying architectural components, offering a holistic strategy accommodating the development of capabilities in LLM-based applications.

% \bibliographystyle{IEEEtran}
% \bibliography{references}

\end{document}